\documentstyle[epsfig]{aipproc}

\begin{document}
\begin{flushright}
\begin{minipage}{5cm}
\begin{flushleft}
\small
\baselineskip = 13pt
YCTP-P19-99\\ hep-ph/9907313 \\
\end{flushleft}
\end{minipage}
\end{flushright}

\title{
Chiral Phase Transition for
\mbox{\boldmath${SU(N)}$} Gauge
Theories}

\author{Francesco Sannino}
\address{Department of Physics, Yale University, New Haven,~CT~06520-8120,~USA}

\maketitle

\begin{abstract}
We describe \cite{SaSc} the chiral phase transition for vector-like
$SU(N)$ gauge theories as a function of the number of quark flavors
$N_f$ by making use of an anomaly-induced effective potential. The
potential depends explicitly on the full $\beta$-function and the
anomalous dimension $\gamma$ of the quark mass operator. By using
this potential we argue that chiral symmetry is restored for
$\gamma <1$. A perturbative computation of $\gamma$ then leads to
an estimate of the critical value $ N_f^c$ for the transition.
\end{abstract}

\section*{Introduction}

The phase structure of strongly coupled gauge field theories as a
function of the number of matter fields $N_f$ is a problem of
general interest. Much has been learned about the phases of
supersymmetric theories in recent years
\cite{Seiberg,Seiberg-Witten,IntSeiberg,Peskin,DiVecchia}. An
equally interesting problem is the phase structure of a
non-supersymmetric $SU(N)$ theory as a function of the number of
fermion fields $N_f$. At low enough values of $N_f$, the chiral
symmetry $SU(N_{f})_L \times SU(N_{f})_R$ is expected to break to
the diagonal subgroup. At some value of $N_f$ less than $11N/2$
(where asymptotic freedom is lost), there will be a phase
transition to a chirally symmetric phase. Whether the transition
takes place at a relatively small value of $N_f$ \cite{mawhinney}
or a larger value remains unknown. The larger value ($N_f / N
\approx 4$) is suggested by studies of the renormalization group
improved gap equation \cite{ATW} and is associated with the
existence of an infrared fixed point. A recent analysis \cite{ASe}
indicates that instanton effects could also trigger chiral symmetry
breaking at comparably large value of $N_f/N$.  Besides being of
theoretical interest, the physics of a chiral transition could have
consequences for electroweak symmetry breaking \cite{AS}, since
near-critical gauge theories provide a natural framework for
walking technicolor theories \cite{ATW2}.

If a phase transition is second order, a useful approach is to find
a tractable model in the same universality class. For chiral
symmetry, a natural order parameter is the $N_f\times N_f$ complex
matrix field $M$ describing mesonic degrees of freedom. If the
meson degrees of freedom are the only ones that develop large
correlation lengths at the phase transition, then the transition
may be studied using an effective Landau-Ginzburg theory.

For the zero-temperature transition as a function of $N_f$, a
similar approach might also be tried. It was suggested in
Ref.\cite{ATW}, however, that while the order parameter vanishes
continuously as $N_f \rightarrow N_f^c$, the transition is not
second order. With the gap equation dominated by an infrared fixed
point of the gauge theory, the transition was argued to be
continuous but infinite order. It has also been noted \cite{sekhar}
that because of the associated long range conformal symmetry, the
masses of all the physical states, not just the scalar mesons are
expected to scale to zero with the order parameter.

We nevertheless suggest that an effective potential using only the
low lying mesonic degrees of freedom might be employed to model at
least some aspects of the zero-temperature chiral phase transition.
The key ingredient is the presence of a new non-analytic potential
term that emerges naturally once the anomaly structure of the
theory is considered. The anomalies also provide a link between
this effective potential term and the underlying gauge theory.

To deduce the anomaly induced effective potential we modify an
effective potential \cite{toy,HSS,Sannino} developed for $N_f< N$,
and apply it to the range $N_f > N$.

We use this potential to discuss the zero-temperature phases of an
$SU(N)$ gauge theory as a function of $N_f$. Assuming that the
transition is governed by an infrared fixed point of the theory, we
deduce that chiral symmetry is restored, together with long-range
conformal symmetry, when $\gamma <1$, where $\gamma$ is the
anomalous dimension of the mass operator. Finally we note that by
using the perturbative expansion of $\gamma $, chiral symmetry is
predicted to be restored above $N_f^c \approx 4N $, in agreement
with a gap equation analysis.

\section*{The Effective Potential}

In this section we construct an effective potential valid to all orders in
the loop expansion and appropriate for the range $N_{f}>N$. The new
ingredients are:

\begin{itemize}
\item[i)]  {Using the full, rather than the one loop, beta function in the
trace anomaly saturation.}

\item[ii)]  {Taking account of the anomalous dimension of the fermion mass
operator.}
\end{itemize}

This anomaly-induced effective potential is based on the QCD trace
and $U_A(1)$ anomalies (see \cite{SaSc} for more details).

We build the potential out of the $N_f\times N_f$ complex meson
matrix $M_i^j$ transforming as the operator $q_i\tilde{q}^j$. So we
assign naive mass dimension 3 to $M_i^j$. The operator $q\tilde{q}$
acquires an anomalous dimension $\gamma $ when quantum corrections
are considered and the full dynamical dimension is thus $3-\gamma$.
To make our effective potential capture the low-energy quantum
dynamics of the underlying theory, we take $ 3-\gamma$ to be the
scaling dimension of $M_i^j$. The anomalous dimension $\gamma$ is
of course a function of the coupling $g$, which in turn depends on
the relevant scale.

To build the final meson potential we only use the chirally
invarian term ${\rm det}M$. This is plausible (see section VII of
Ref.~{\cite {GJJS}}) and would correspond to the ''holonomic''
structure which emerges if the potential is considered to arise
(\cite{toy,HSS,Sannino}) from broken super QCD. In the same spirit
we take det$M$ to have the scaling dimension $(3-\gamma )N_{f}$.

The potential term we find is \cite{SaSc}
\begin{equation}
V=-C \Lambda^{4}\left[ \frac{\Lambda ^{3N_f}} {{\rm det}M}\right] ^
{\frac 4{f(g)}}+{\rm h.c.} \ ,  \label{ass}
\end{equation}
where $C$ is related to $A$ via:
\begin{equation}
C=\frac {f(g)}{4e}\exp \left[ {\frac{4A} {f(g)}}\right] \ ,
\label{coefficient}
\end{equation}
and
\begin{equation}
f(g)=-\frac{\beta(g)}{g^3}16\pi^2 - (3-\gamma) N_f \ .
\label{function}
\end{equation}
 Finally we integrate out the $\eta^{\prime}$ field, which can
be isolated by setting
\begin{equation}
{\rm det}M=|{\rm det}M|e^{i\phi }\ ,  \label{prime}
\end{equation}
where $\phi \propto \eta ^{\prime }$. This is done anticipating
that the $\eta^{\prime}$ will be heavy with respect to the
intrinsic scale of the theory and the other mesonic degrees of
freedom. Now using Eq.~(\ref{prime}) we derive the field equation
$\phi=0$ which leads to the final potential
\begin{equation}
V=-2C \Lambda^{4}\left[\frac{\Lambda^{3N_f}} {|{\rm det} M|}\right]^{\frac{4%
}{f(g)}} \ .  \label{inter}
\end{equation}
The shape of this potential is determined by the function $f(g)$
Eq.~(\ref {function}).

Our interest here is in the range $N < N_f < (11/2)N$ where the
chiral phase transition is expected to occur. For $N_f$ close to
$(11/2)N$, a weak infrared fixed point will occur. The $\beta$
function will be negative and small at all scales and $\gamma$ will
also be small. Thus $f(g)$ will be negative. As $N_f$ is reduced,
the fixed point coupling increases as does $\gamma$.
 However in the range of interest $f(g)$
will remain negative ($(3-\gamma)N_f >
-(\beta(g)/g^{3})16\pi^2$). The potential in Eq.~(\ref{inter}) may then be
written as
\begin{equation}
V=+2|C|\Lambda^{4}\left[ \frac{|{\rm det}M|} {\Lambda ^{3N_f}}
\right]^{\frac 4{\frac{\beta (g)}
{g^3}16\pi ^2+(3-\gamma )N_f}}\ .  \label{grandeNf}
\end{equation}
It is positive definite and vanishes with the field $|{\rm det}M|$.

\subsection*{The Chiral Phase Transition}

To study the chiral phase transition, we need the combined effective
potential
\begin{equation}
V_{tot}=V + V_I \,
\end{equation}
where $V_I$ is a generic potential term not associated with the
anomalies. It is instructive, however, to investigate first the
extremum properties of the anomaly term (Eq.~(\ref{grandeNf})).
Assuming the standard pattern for chiral symmetry breaking
$SU_R(N_f)\times SU_L (N_f)\rightarrow SU_V (N_f)$, $M^i_j$ may be
taken to be the order parameter for the transition. For purposes of
this discussion, we restrict attention to the vacuum value of $
M^i_j$, which can be rotated into the form $M^i_j=\delta^i_j \rho$,
where $\rho \geq 0$ is the modulus. Substituting the previous
expression in the anomaly induced effective potential gives the
following expression:
\begin{equation}
V=+2|C|\Lambda^{4} \left[\frac{\rho}{\Lambda^{3}}
\right]^{\frac{4N_f}{\frac{
\beta(g)}{g^3}16 \pi^2 + (3 - \gamma)N_f }} \ .  \label{rhopotential}
\end{equation}
Recall that ($(3-\gamma)N_f > -(\beta(g)/g^{3})16\pi^2$) in the
range of interest. The first derivative $\partial V/\partial \rho$
vanishes at $\rho=0 $ provided that $\frac{4N_f}{\frac{\beta
(g)}{g^3}16\pi ^2+(3-\gamma )N_f}>1$, a condition that is clearly
satisfied. The second derivative,
\begin{equation}
\frac{\partial ^2V}{\partial \rho ^2}\propto \rho ^{\left[\frac{4N_f}{\frac{
\beta(g)} {g^3}16\pi ^2+(3-\gamma )N_f}-2 \right]} \ ,  \label{2nd}
\end{equation}
also vanishes at $\rho =0$ if the exponent in Eq.~(\ref{2nd}) is
positive. The second derivative at $\rho =0$ is a positive constant
when the exponent vanishes, and it is $+ \infty$ for
\begin{equation}
\frac{4N_f}{\frac{\beta (g)}{g^3}16\pi ^2+(3-\gamma )N_f}-2<0 \ .
\label{criticality}
\end{equation}
The curvature of $V_{tot}$ at the origin is given by the sum of the
two terms $\frac{\partial ^2V}{\partial \rho ^2}$ and
$\frac{\partial^2V_I}{\partial \rho ^2}$, evaluated at $\rho = 0$.

To proceed further, we assume that the phase transition is governed by an
infrared stable fixed point of the gauge theory. We thus set $\beta (g)=0$.
The curvature of V at the origin is then $0$ for $\gamma >1$, finite and
positive for $\gamma =1$, and $+\infty $ for $\gamma <1$. The value of $%
\gamma $ depends on the fixed point coupling, which in turn depends on $N_{f}
$. As $N_{f}$ is reduced from $(11/2)N$, the fixed point coupling increases
from $0$, as does $\gamma $. Assuming that $\gamma $ remains monotonic in $%
N_{f}$, growing to $1$ and beyond as $N_{f}$ decreases, there will be some
critical value $N_{f}^{c}$ below which $\frac{\partial ^{2}V}{\partial \rho
^{2}}$ vanishes at the origin. The curvature of $V_{tot}$ will then be
dominated by the curvature of $V_{I}$ at the origin. For $N_{f}=N_{f}^{c}$,
there will be a finite positive contribution to the curvature from the
anomaly-induced potential. For $N_{f}>N_{f}^{c}$ ($\gamma <1$), $V$
possesses an infinite positive curvature at the origin, suggesting that
chiral symmetry is necessarily restored. We will here take the condition $%
\gamma =1$ to mark the boundary between the broken and symmetric phases, and
explore its consequences. This condition was suggested in Ref.
\cite{CG}, based on other considerations.

We next investigate the behavior of the theory near the transition
by combining the above behavior with a simple model of the
additional, non-anomalous potential $V_I$. We continue to focus
only on the modulus $\rho $ and take the potential to be a
traditional Ginzburg-Landau mass term, with the squared mass
changing from positive to negative as $\gamma - 1$ goes from
negative to positive: $\left( 1 - \gamma \right) \Lambda ^{-2}\rho
^2$.
Additional, stabilizing terms, such as a $\rho^4$ term, could be added but
will not affect the qualitative conclusions. The full potential is then
\begin{equation}
V_{tot}=2|C|\Lambda^{4}(\frac{\rho}{\Lambda^3})^{\frac{4}{3-\gamma}} -
\left(\gamma - 1\right)\Lambda ^{-2} \rho^2 \ .  \label{potenzialetto}
\end{equation}
{}For $\gamma >1$ (but $<3$), the first term stabilizes the potential for
large $\rho$, and the potential is minimized at
\begin{equation}
<\rho >=\Lambda^3 \left[\frac{\gamma - 1} {2|C|}\right]^{\frac{1}{\gamma-1}}
\ ,  \label{lexponent}
\end{equation}
in the limit $\gamma \rightarrow 1$. It describes an infinite order
phase transition as $\gamma
\rightarrow 1$ , in qualitative agreement with the gap equation
studies. This behavior would not be changed by the addition of
higher power terms ($\rho^4, \rho^6,
...$) to the potential.

The curvature of the potential Eq.~(\ref{potenzialetto}) at the
minimum describes a mass associated with the field $\rho $. To
interpret this mass physically, one should construct the kinetic
energy term associated with this field (at least to determine its
behavior as a function of $\gamma -1$ ). We hence rescale $\rho $
to a field $\sigma $ via $\displaystyle{\rho
=\sigma ^{3-\gamma }\Lambda ^{\gamma }}$ with $\sigma $ possessing a
conventional kinetic term $-\frac{1}{2}(\partial ^{\mu }\sigma
)^{2}$.
This then leads to the following result for the physical
mass $M_{\sigma }$ and $<\sigma >$
\begin{equation}
<\sigma >\simeq \left[ \frac{\gamma -1}{2|C|}\right]
^{\frac{1}{2(\gamma -1)} }\Lambda \ ,\qquad M_{\sigma }\simeq
2\sqrt{6}|C|\left[ \frac{\gamma -1}{2|C|}\right]
^{\frac{1}{2(\gamma -1)}}\Lambda \ .
\end{equation}

Likewise, in the presence of the quark mass term we have (see
\cite{SaSc} for details)
\begin{equation}
\left[ <\sigma >\right] _{\gamma =1}\simeq \left[ \frac{mN_{f}\Lambda }{2|C|}
\right] ^{\frac{1}{2}}\ ,\qquad \left[ M_{\sigma }\right] _{\gamma =1}\simeq
2\left[ 2mN_{f}\Lambda \right] ^{\frac{1}{2}}\ .
\end{equation}
Thus the order parameter $\sigma $ for $\gamma =1$ vanishes according to the
power $1/2$ with the quark mass in contrast with an ordinary second order
phase transition where the order parameter is expected to vanish according
to the power $1/3$.

Finally we note an important distinction between our effective
potential describing an infinite order transition and the
Ginzburg-Landau potential describing a second order transition. The
latter may be used in both the symmetric and broken phases,
describing light scalar degrees of freedom as the transition is
approached from either side. Our potential develops infinite
curvature at the origin in the symmetric phase, indicating that no
light scalar degrees of freedom are formed as the transition is
approached from that side. This is in agreement with the
conclusions of Ref. \cite{ARTW}, indicating that as one crosses to
the symmetric phase, mesons melt into quarks and gluons and hence
the physics is described via only the underlying degrees of
freedom. The present effective Lagrangian formalism for describing
the chiral/conformal phase transition is close in spirit to the one
developed in Ref.\cite{MY}.

By perturbatively (see \cite{SaSc}) saturating at two loops the
condition $\gamma<1$ at the fixed point value of the coupling
constant leads to the conclusion that chiral symmetry is restored
for $N_f>N_f^c \simeq 3.9 N$.

\section*{Conclusions}
 We have explored the chiral phase transition for vector-like $SU(N)$ gauge
theories as a function of the number of flavors $N_f$ via an
anomaly induced effective potential. The effective potential was
constructed by saturating the trace and axial anomalies. It depends
on the full beta function and anomalous dimension of the quark-mass
operator. We showed that the anomaly induced effective potential
for $N_{f}>N$ is positive definite and vanishes with the field
$M_{i}^{j}$. We then investigated the stability of the potential at
the origin, and discovered that the second derivative is positive
and divergent when the underlying $\beta$ function and the
anomalous dimension of the quark-mass operator satisfy the relation
of Eq.~(\ref{criticality}). We took this to be the signal for
chiral restoration. With conformal symmetry being restored along
with chiral symmetry (due to the $\beta $ function vanishing at an
infrared fixed point), the criticality relation becomes a
constraint on the anomalous dimension of the quark-mass operator:
\begin{equation}
\gamma <1\ .
\end{equation}
To convert this inequality into a condition for a critical number
of flavors, we used the perturbative expansion of the anomalous
dimension evaluated at the fixed point, deducing that chiral
symmetry is restored for $N_{f}\simeq 4N$, in agreement with gap
equation studies.

The core of this talk is the proposal that the chiral/conformal
phase transition, suggested by gap equation studies to be
continuous and infinite order, may be described by an effective
potential whose form is dictated by the trace and axial anomalies
of the underlying $SU(N)$ gauge theory.
\acknowledgments
I am very happy to thank Joseph Schechter for sharing the work on
which this talk is based and Thomas Appelquist for enlightening
discussions. The work has been partially supported by the US DOE
under contract DE-FG-02-92ER-40704.

\end{document}